# *Characteristics of the Sculptured Cu Thin Films and Their Optical Properties as a function of deposition rate*

H. Savaloni[*,1], F. Babaei[1], S. Song[2], F. Placido[2]

*1) Department of Physics, University of Tehran, North-Kargar Street, Tehran, Iran.*
*2) Thin Film Center, The University of The West of Scotland, High Street, Paisley, UK.*
*\*) Corresponding author: Tel: +98 21 6635776; Fax: +98 21 88004781; Email: savaloni@khayam.ut.ac.ir*

**Abstract**

Sculptured copper thin films were deposited on glass substrates, using different deposition rates. The nano-structure and morphology of the films were obtained, using X-ray diffraction (XRD), atomic force microscopy (AFM) and scanning electron microscopy (SEM). Their optical properties were measured by spectrophotometry in the spectral range of 340-850 nm. The Scot package was used for modeling the film structure and fitting the calculated optical transmittance results to the experimental data and obtaining, both real and imaginary refractive indices, film thickness and fraction of metal inclusion in the film structure.

*Keywords: Chiral sculptured thin films; Bruggeman formalism*

**1. Introduction**

Sculptured thin films (STFs) are columnar (range between 1 to 100 nm) thin films deposited on substrate with controlled azimuthal rotation, $\phi$, and tilt angle, $\alpha$, by a method known as glancing angle deposition (GLAD) [1-3].

The understanding and modeling [4-6] of sculptured films deposited on tilted substrates has become increasingly important as their applications encompass various disciplines: photonics [7-10], liquid crystal display technology [11], magnetic media information storage [12], organic or inorganic sensors [13], and energy storage technology [2], among others.

In this paper we report on the nano-structural and optical properties of sculptured copper thin films produced using different deposition rates.

## 2. Experimental details

The copper sculptured thin films were deposited on glass substrates (microscope slide) by resistive evaporation from tungsten boats at room temperature. The purity of copper was 99.99%. An Edwards (Edwards E19 A3) coating plant with a base pressure of $2 \times 10^{-6}$ mbar was used. The deposition angle was fixed at 75 degrees and a substrate azimuthal rotation speed of 0.1 RPM was chosen. Fig. 1 shows the schematic of the evaporation system showing substrate position and rotation for sculptured chiral/pillar thin film growth. The movement of the stepper motor and its speed of revolution as well as facility for dividing each revolution to different sectors are controlled through interface to a computer in which the related software is written and installed. All these are domestic made.

Prior to deposition, all glass substrates were ultrasonically cleaned in heated acetone then ethanol. The surface roughness of the substrates was measured by a Talysurf profilometer and AFM and the rms substrate roughness $R_q$ obtained using these methods was 0.3 nm and 0.9 nm, respectively.

In order to provide a point source for geometrical considerations, a plate of tungsten with a 6 mm diameter hole in the middle was used as a mask on top of the evaporation boat. The deposition process was repeated a few times and the reproducibility of the results was confirmed. The deposition rate was measured by a quartze crystal deposition rate controller (Sigma Instruments, SQM-160, USA) positioned close to the substrate and at the almost the same azimuthal angle as that of the substrate with considering the masking effect of the quartze crystal (the mask was removed in order to have the crystal exposed to the same incident beam (vapour direction) as that of the substrate.

The film thicknesses and column shapes and sizes were measured by field emission electron microscope (FESEM) .The SEM samples were coated with a very thin layer of gold to prevent the charging effect.

The nanostructure of these films was obtained using a Siemens D500 x-ray Diffractometer (CuK$_\alpha$ radiation; 40 kV, 30 mA) with a step size of 0.02° and count time of 1 s per step, while the surface physical morphology and roughness was obtained by means of AFM analysis. The transmittance spectra of the samples were obtained using a double beam spectrophotometer in the spectral range of (340-850 nm) corresponding to the energy range of (3.65-1.46 eV).

## 3. Results
*Crystallographic and nano-structure of the films*

Figure 2 shows the XRD patterns of the as deposited Cu chiral/pillar sculptured thin films produced with deposition rates of 0.2, 0.5 and 2.5 Ås$^{-1}$, respectively. It can be observed that as expected and reported in a number of literatures [e.g, 14-18] the diffraction pattern obtained for all these films exhibits almost an amorphous structure with a weak diffraction peak at ($2\theta = 43.72°$) peculiar to the Cu(111) diffraction line of Cu.

The 2D AFM images of the films produced in this work are arranged according to their deposition rates in Figs. 3(a-c). It is worthwhile to emphasis that in particular the 2D AFM images (Figs. 3(a-c)) show very intriguing features which are entirely different from those reported for columnar thin films produced without rotation of the substrate about its azimuthal axis and those produced at normal incidence. Although we have had no intention of growing nano-flowers [19] for which one has to vary the speed of rotation within each revolution (i.e., each revolution has to be divided into 2N sectors, where N is the symmetry of desired shape of the columns (nano-flower)), but we may claim that the AFM images (Figs. 3(a-c)) clearly show the formation of the nano-flowers though their symmetry is not as regular as those reported by Zhao et al [19]. In Figs. 3(a´-c´) selected parts of Figs. 3(a-c) (shown using boxes on Figs. 3(a-c)) are presented which clearly show the above mentioned nano-flowers.

The grain size distributions of these films were obtained from the 2D AFM images. In Table 1 the grain size, surface roughness (obtained from AFM analysis) and both surface fraction of metallic inclusions obtained from the analysis of the AFM images and those obtained from the optical analysis of the samples (discussed in section 3.2) are given. It can be seen that the surface roughness decreases with increasing the deposition rate. This can be the result of increased number of buried atoms with increasing the deposition rate (the number of diffused atoms decreases with increasing the deposition rate, because of the bombardment of atoms with the following atoms), which produces smaller grains (densely packed fibrous structure, seen in the SEM photographs of these films Figs. 4(a-c)). Hence it is expected to obtain higher packing density (fraction of inclusions in the film structure, which is consistent with the results obtained from the analysis of the 2D images of these samples (column 7 of Table 1) as well as the results obtained from the structural model designed for obtaining the optical data for these samples (column 8 of Table 1, discussed in section 3.2.

Table 1. Structural data of the chiral/pillar sculptured Cu thin films obtained using different techniques

| Sample | Deposition rate ($Ås^{-1}$) | Grain(Min) | Grain(Ave) | Grain (Max) | RMS Surface roughness (AFM results)(nm) | Surface fraction (metallic inclusion) (AFM) | Fraction of metallic inclusions in the film structure (OPTICAL) |
|---|---|---|---|---|---|---|---|
| **6a** | 0.2 | 28nm | 76nm | 179nm | 15.416 | 0.72 | 0.73 |
| **6b** | 0.5 | 26nm | 65nm | 135nm | 15.254 | 0.76 | 0.86 |
| **6c** | 2.5 | 21nm | 62nm | 120nm | 14.733 | 0.78 | 0.92 |

The SEM images of these films (Figs. 4 (a-c)) also clearly show that by increasing the deposition rate the fraction of voids decreases and denser films are formed. This is further confirmed by fitting the optical transmission data, using SCOT program, in which the structure of the film should be modelled (see section 3.2).

It can also be seen that the surface of the film becomes more uniform, which is consistent with the roughness data obtained from the analysis of the AFM images. The formation of two layers during the process of the deposition of sculptured thin films, namely bottom layer and the top layer can also be observed in these pictures [20]. Due to thinness (and perhaps due to high deposition rates used) of these samples ( ~ 90 nm) there is a difficulty to observe the pitches of the chiral structure, although in Fig. 4(b) a few columns which are pointed by arrows clearly show three pitches formed in this sample.

*3.2. Optical properties*

The optical properties of the samples were measured by spectrophotometry n the spectral range of 340-850 nm. The Scot package was used for modeling the film structure and fitting the calculated optical transmittance and ellipsometery results to the experimental data and obtaining, real and imaginary refractive indices, film thickness, using the extended Drude model and fraction of metal inclusion in the film structure, using Bruggeman effective media approximation [21,22].

It was found that the best structural model for fitting the optical data is a structure with two layers, namely bottom layer (metallic condensed layer) and top layer (mixture of Cu chiral/pillar inclusions and voids) as is also reported by researchers in this field (e.g., [23]) and is observed in the SEM micrographs of this work (Figs. 5(a-c)).

For the bottom layer we used the metal dispersion model (i.e., Extended Drude Model plus Kim oscillators [23] and the dielectric background). The classical Drude model works with a damping constant which does not depend on frequency. This is a good approximation in most cases. However, there are situations where the damping of the free carriers exhibits a characteristic dependence on frequency. A simple choice of the damping term is:

$$\chi_{Drude}(\nu) = -\frac{\Omega_p^2}{\nu^2 + i\nu\Omega_\tau} \quad \text{with} \quad \Omega_p^2 = \frac{ne^2}{\varepsilon_0 m}$$

$$\Omega_\tau(\nu) = \Omega_{\tau,Low} - \frac{\Omega_{\tau,Low} - \Omega_{\tau,High}}{\pi}\left[\arctan\left(\frac{\nu - \Omega_{\tau,Crossover}}{\Omega_{\tau,Width}}\right) + \frac{\pi}{2}\right]$$

where, $\Omega_p$ is the plasma frequency, $\Omega_{\tau,Low}$ is the damping constant at low frequency, $\Omega_{\tau,High}$ is damping constant at high frequency, $\Omega_{\tau,Crossover}$ is the cross over frequency, centre of the transition region, $\Omega_{\tau,Width}$ is the width of the transition region, $n$ is the charge carrier density, $m$ is the effective mass and $e$ is the elementary charge. The function for the damping constant is chosen to change smoothly from a constant at low frequencies to another constant level at high frequency. The transition region is defined by the crossover frequency and the width parameter. The parameters used for the analysis of the samples in this work were set at the following values: $\Omega_p = 1000$, $\Omega_{\tau,Low} = 2000$, $\Omega_{\tau,High} = 100$, $\Omega_{\tau,Crossover} = 3000$, $\Omega_{\tau,Width} = 1000$.

The frequency dependence of the damping constant was achieved by allowing a continuous shift of the line shape between a Gaussian and Lorentzian profile using the vibrational modes suggested by Kim et al [23]:

$$\chi_{Kimoscillator} = \frac{\Omega_p^2}{\Omega_{IO}^2 - \nu^2 - i\nu\tau(\nu)} \quad \text{with} \quad \tau(\nu) = \Omega_\tau \exp\left(-\frac{1}{1+\sigma^2}\left(\frac{\nu - \Omega_{TO}}{\Omega_\tau}\right)^2\right)$$

The constant $\sigma$ is called Gauss-Lorentz switch. Like almost all fit parameters it may vary between 0 and infinity. For $\sigma = 0$ a Gaussian line shape is achieved. Large values of $\sigma$ (larger than 5) lead to a Lorentzian line shape.

The top layer (mixed layer as metallic inclusions and voids) was modeled, using the Bruggeman effective medium approximation [21,22] as:

$$(1-f)\frac{\varepsilon_m - \varepsilon_{eff}}{\varepsilon_m + 2\varepsilon_{eff}} + f\frac{\varepsilon_p - \varepsilon_{eff}}{\varepsilon_p + 2\varepsilon_{eff}} = 0$$

in which $\varepsilon_m$, $\varepsilon_p$, and $\varepsilon_{eff}$ are the dielectric constants of the material (e.g., metal), void, and the effective medium, respectively, and $f$ is the fraction of material inclusion in the body of the medium [21,22].

The values of $f$ obtained from the fitting procedure of the experimental transmissions of the samples to this model using the Scot programme are given in column 8 of Table 1. These results are in good agreement with those obtained from the analysis of the AFM images which are given in column 7 of Table 1. Hence, confirming our modeling approach, though the latter is related to the surface structure and the former one represents the structural (bulk) data.

The other feature which may influence the optical results is the film surface roughness and morphology. Light is scattered by the rough surface and the light loss need to be taken into account. An exponential relation between light loss, $LL_{surfrough}$ and wavelength is used for this layer based on the assumptions of Gaussian distribution of roughness and coherent light scattering, as:

$$LL_{surfrough} \approx \exp\left\{-\left[\frac{3\pi c_1 c_2}{\lambda(nm)}\right]^2\right\}$$

where $c_1$ is the film surface roughness and $c_2$ is the air refractive index (unity).

Figures 5(a-c) show the measured transmission results and their fitting quality to the model. In Figs. 6(a and b) the results obtained for the real and imaginary parts of the refractive index of the samples are compared, respectively. It can be observed that

With regard to void fraction calculation, usually an over-layer composed of oxide and surface micro-roughness is also considered. In addition one may also point out that the effect of oxide formation on crystalline samples is more evident in vuv region. In the infrared, the oxides tend to be transparent and have negligible effects. In the visible, the effects of a transparent oxide become greater, and in the ultraviolet, where the oxide is absorbing, qualitative and quantitative effects can occur. In general, the effect of a thin transparent oxide layer is to reduce the reflectivity [24]. Since in non of the results of $\varepsilon_2$ for Cu chiral/pillae sculptured thin films (Figs.7) a shift in peak energy can be seen, therefore we may conclude that an over-layer is not formed on our films, while the shift in the height of the $\varepsilon_2$ spectra is an indication of the void content in the films [21].

The results for the parameters used in fitting of the model to the experimental transmissions are summarized in Table 2.

## 4. Conclusions

The influence of deposition rate on the nano-structure and optical properties of sculptured copper thin films are investigated using XRD, AFM, SEM and spectrophotometery. The 2D AFM images showed structures resembling the formation of nano-flowers, though they do not have regular symmetries as those films designed for growing nano-flowers. The film surface void fraction obtained using the AFM images and those obtained from optical data fitting, using the Scot Code were consistent, while the void fraction decreased with increasing the deposition rate. This caused a shift in the height of the $\varepsilon_2$ results.

**Acknowledgements**

This work was carried out with the support of the University of Tehran and

**Figure captions**

Figure 1. Schematic of the GLAD process.

Figure 2. XRD patterns of the sculptured Cu thin films produced with different deposition rates of a) 0.2 nms$^{-1}$, b) 0.5 nms$^{-1}$, c) 2.5 nms$^{-1}$, with an azimutal rotation speed of o.1 RPM.

Figure 3. AFM images of the sculptured Cu thin films produced with different deposition rates of a) 0.2 nms$^{-1}$, b) 0.5 nms$^{-1}$, c) 2.5 nms$^{-1}$, with an azimutal rotation speed of o.1 RPM. a', b' and c' are selected parts of a, b, and c, which show typical nano-flowers grown on these films.

Figure 4. SEM Electron Micrographs of the surface and cross section of sculptured Cu thin film, produced with different deposition rates of a) 0.2 nms$^{-1}$, b) 0.5 nms$^{-1}$, c) 2.5 nms$^{-1}$, with an azimutal rotation speed of o.1 RPM.

Figure 5. Transmission spectra of the sculptured Cu thin film, produced with different deposition rates of a) 0.2 nms$^{-1}$, b) 0.5 nms$^{-1}$, c) 2.5 nms$^{-1}$, with an azimutal rotation speed of o.1 RPM.

Figure 6. Optical functions of the sculptured Cu thin film, produced with different deposition rates of a) 0.2 nms$^{-1}$, b) 0.5 nms$^{-1}$, c) 2.5 nms$^{-1}$, with an azimutal rotation speed of o.1 RPM.

Figure 7. The imaginary part of the dielectric function of the sculptured Cu thin films, produced with different deposition rates of a) 0.2 nms$^{-1}$, b) 0.5 nms$^{-1}$, c) 2.5 nms$^{-1}$, with an azimutal rotation speed of o.1 RPM.

Table 2. Fitting parameters used in the SCOT Code

| | #a | #b | #c |
|---|---|---|---|
| Plasma frequency | 46498.6875 | 46882.7891 | 46376.9219 |
| Low damping | 25507.3672 | 27078.1836 | 12748.6592 |
| High damping | 0.0182 | 0.0280 | 6.1631 |
| crossover | 16334.6455 | 16193.6377 | 16753.2637 |
| width | 978.5741 | 810.3342 | 521.5281 |
| Resonance frequency | 20293.1035 | 20229.6914 | 19918.6230 |
| Oscillator strength | 12787.1230 | 13049.8730 | 14800.8535 |
| Damping | 5779.4844 | 5830.0806 | 6723.3193 |
| Gauss-Lorentz Switch | 0.0000 | 0.0000 | 0.8485 |
| Resonance frequency | 28382.3828 | 28327.0703 | 29983.5820 |
| Oscillator strength | 41137.0391 | 39495.3047 | 49220.0625 |
| Damping | 18418.8945 | 18833.2773 | 24754.3438 |
| Gauss-Lorentz Switch | 0.0048 | 0.0073 | 0.2598 |
| Resonance frequency | 14482.9580 | 14508.0762 | 14417.2441 |
| Oscillator strength | 11900.1914 | 13223.6113 | 12739.3320 |
| Damping | 5151.3330 | 4642.9385 | 5119.3545 |
| Gauss-Lorentz Switch | 0.0049 | 0.0081 | 0.0002 |
| Real part, dielectric background | 6.2316 | 6.3967 | 7.3418 |

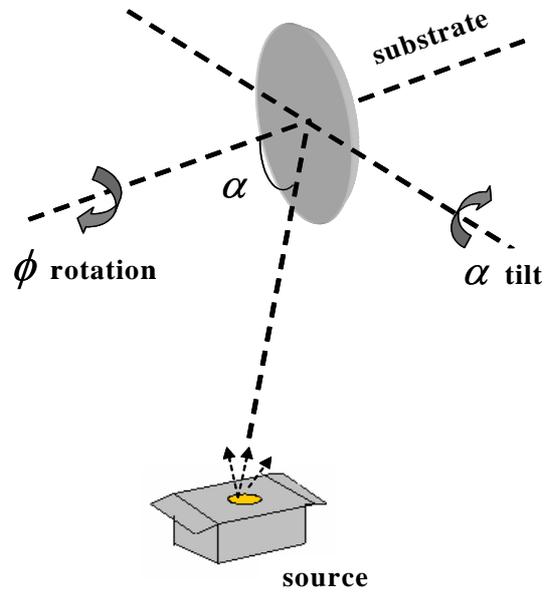

**Fig 1. Savaloni et al.**

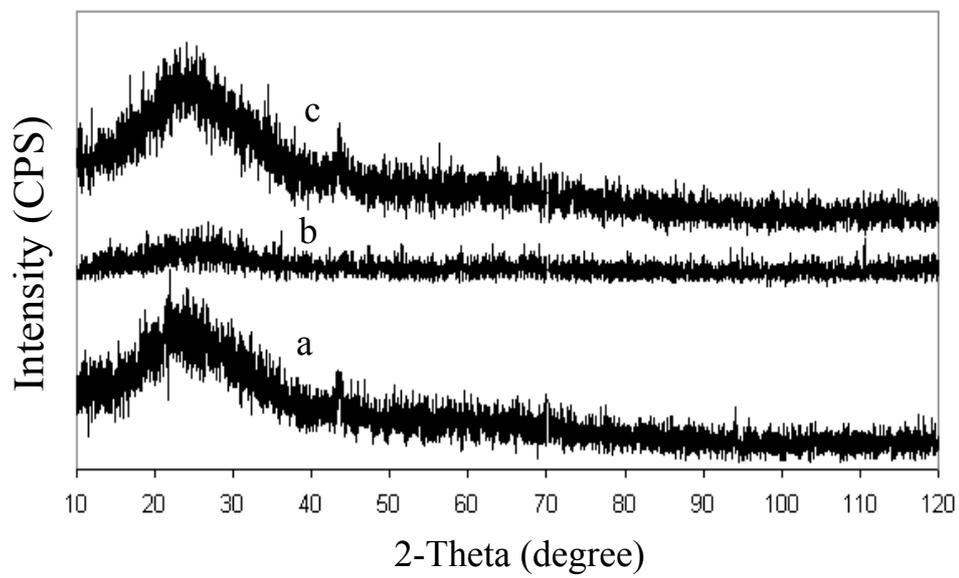

**Fig 2. Savaloni et al.**

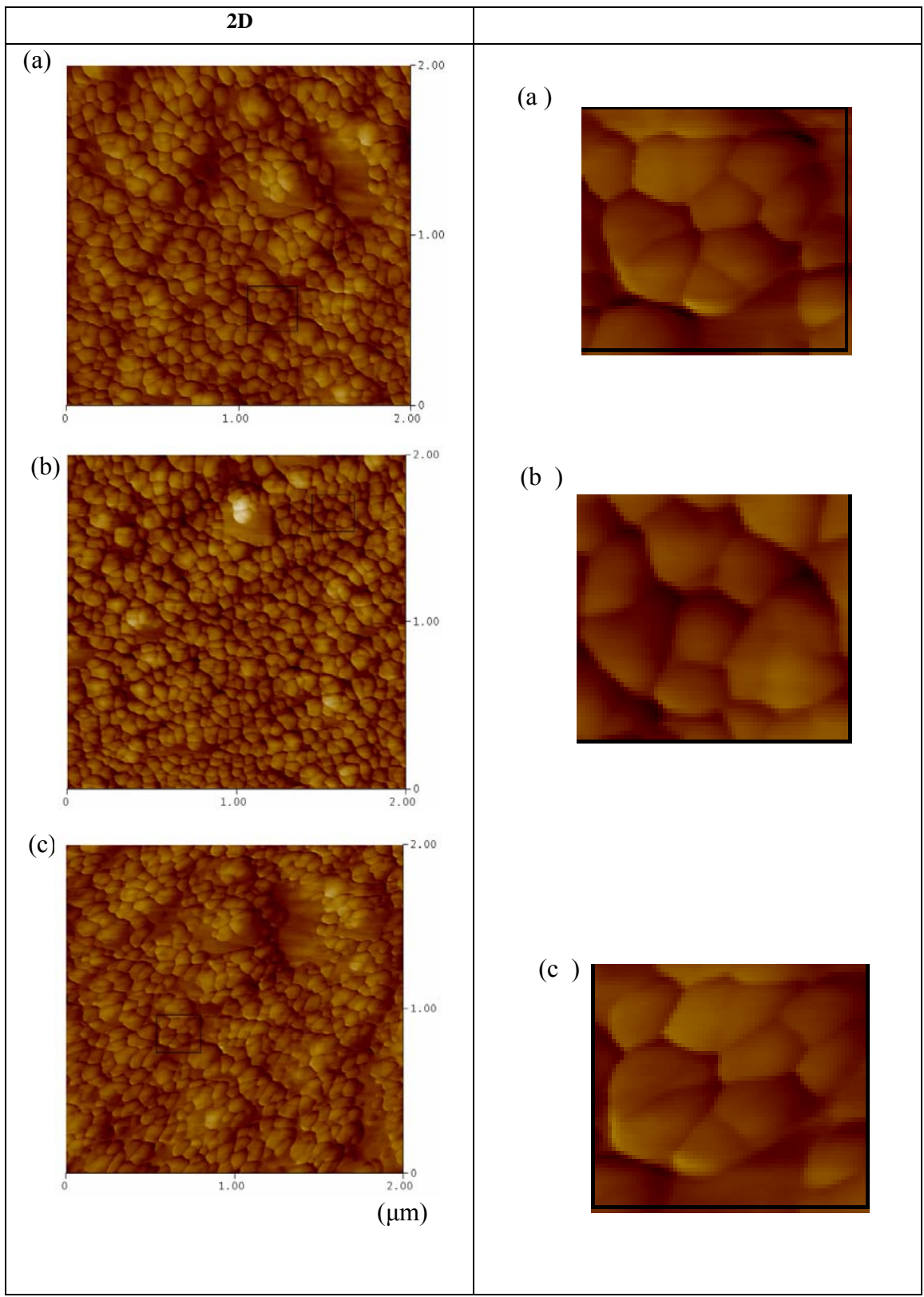

**Fig 3. Savaloni et al.**

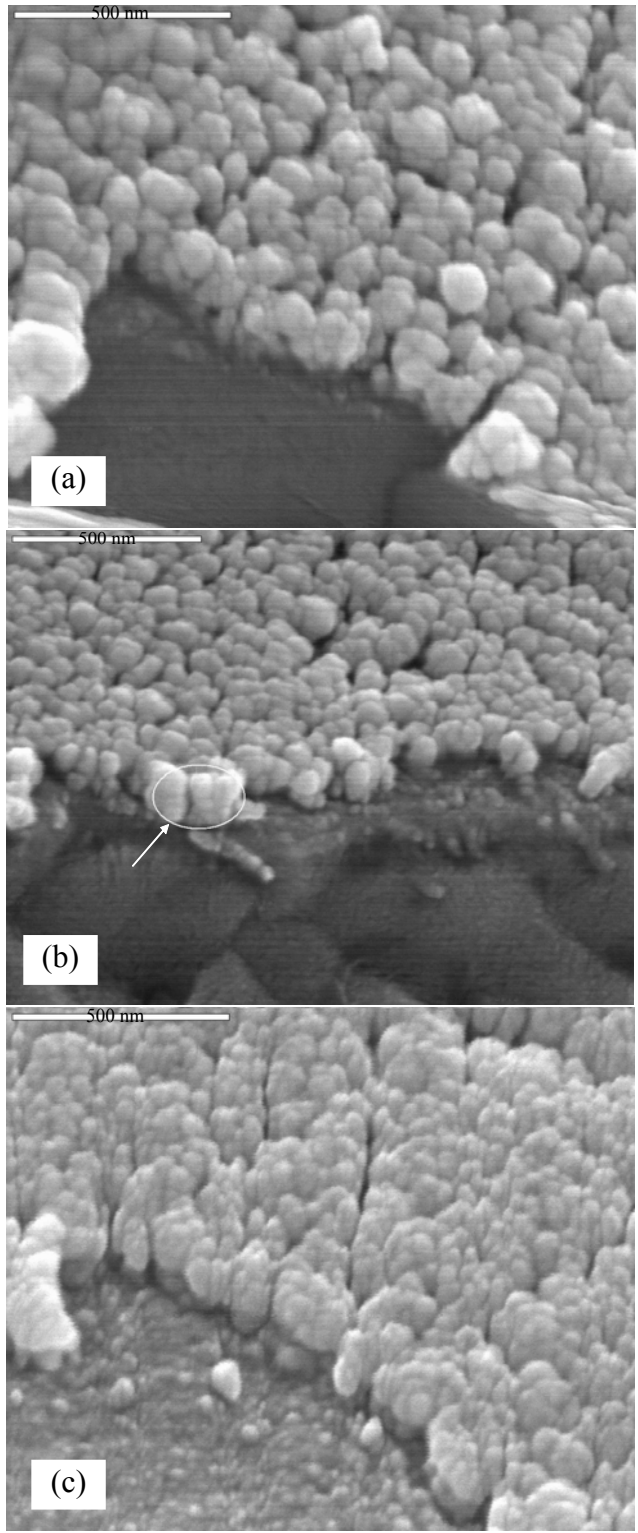

**Fig 4. Savaloni et al.**

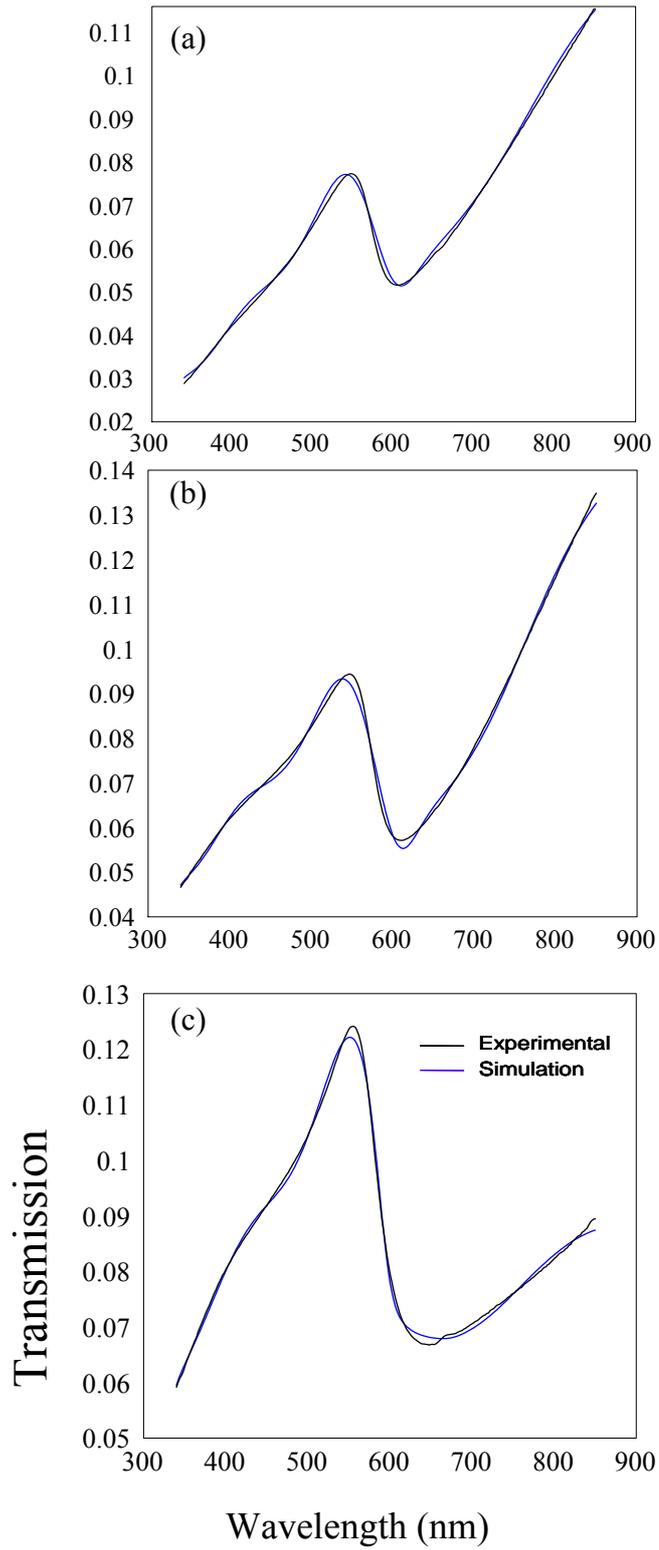

**Fig 5. Savaloni et al.**

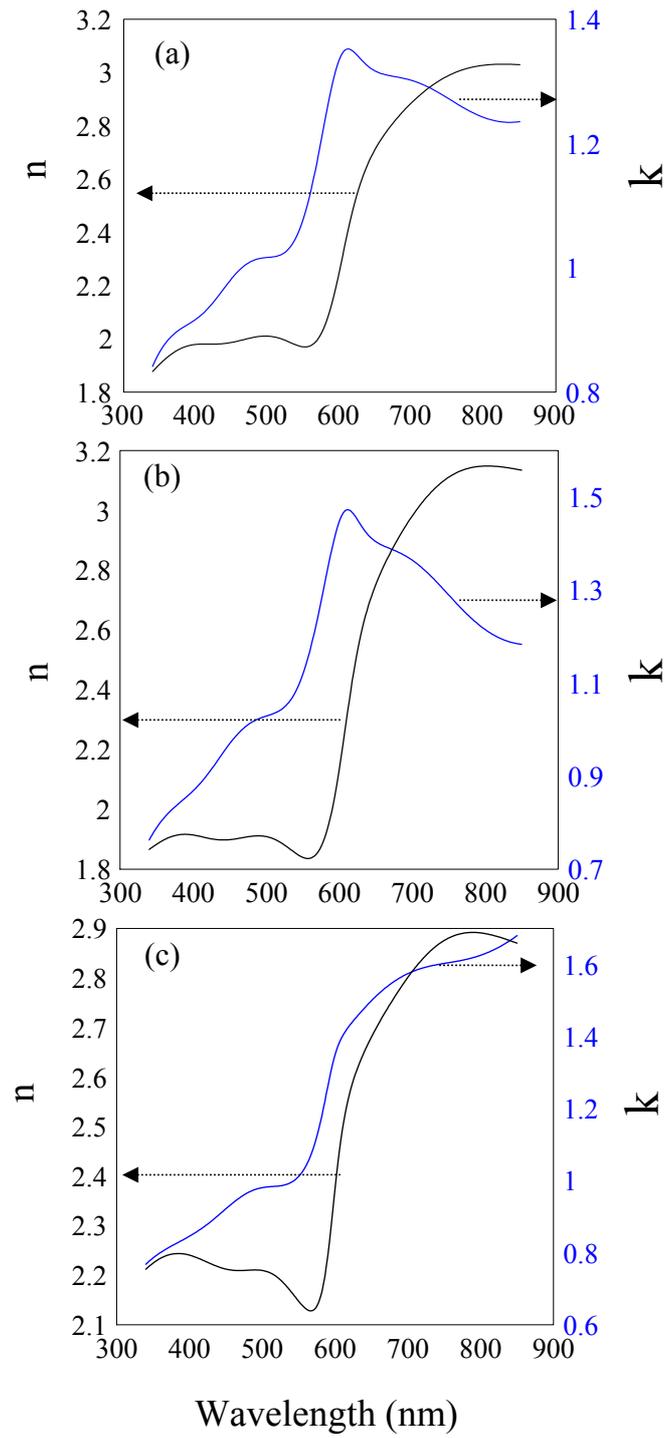

**Fig 6. Savaloni et al.**

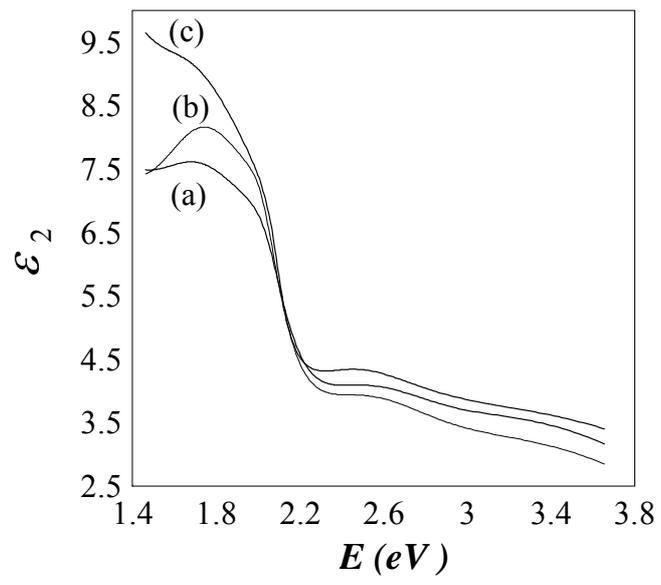

**Fig 7. Savaloni et al.**